\documentclass{article}
\usepackage{spconf,amsmath,epsfig}

\usepackage{lineno}
\usepackage{graphicx}
\usepackage[usenames,dvipsnames,svgnames,table]{xcolor}
\usepackage[utf8x]{inputenc}
\usepackage{amsmath}\usepackage{amssymb}
\usepackage{amsthm}
\usepackage{color}
\usepackage{multirow}
\usepackage{booktabs}
\usepackage{framed}
\usepackage{epstopdf}
\usepackage{url}
\usepackage{ucs}
\usepackage{soul}
\usepackage{breqn}
\usepackage{cite}

\usepackage{hyperref}
\hypersetup{
    unicode=false,          
    pdftoolbar=true,        
    pdfmenubar=true,        
    pdffitwindow=false,     
    pdfstartview={FitH},    
    pdftitle={},    
    pdfauthor={},     
    pdfsubject={},   
    pdfcreator={},   
    pdfproducer={}, 
    pdfkeywords={Landsat}{Fusion}{Reflectance}{MODIS}{Interpolation}, 
    pdfnewwindow=true,      
    colorlinks=true,       
    linkcolor={blue},          
    citecolor=blue,        
    filecolor=blue,      
    urlcolor=blue           
}

\title{Interpolation and gap filling of Landsat reflectance time series}
\name{\'Alvaro Moreno-Mart\'inez$^{1,3}$, Marco Maneta$^2$, Gustau Camps-Valls$^3$, Luca Martino$^{3,4}$}
{Nathaniel Robinson$^1$, Brady Allred$^1$ and Steven W Running$^1$
\thanks{This research was financially supported by the NASA Earth Observing System MODIS project (grant NNX08AG87A) and the European Research Council (ERC) Consolidator Grant SEDAL (Statistical Learning for Earth Observation Data Analysis) project under Grant Agreement 647423.}}
\address{$^1$Numerical Terradynamic Simulation Group (NTSG),  University of Montana, USA,\\
$^2$Department of Geosciences,  University of Montana, USA,\\
$^3$Image Processing Laboratory (IPL), Universitat de Val\`encia, Spain,\\
$^4$Dep. Signal Processing and Communications, Universidad Carlos III de Madrid, Spain}

\begin{document}
%
\maketitle
\begin{abstract}
Products derived from a single multispectral sensor are hampered by a limited spatial, spectral or temporal resolutions. Image fusion in general and downscaling/blending in particular allow to combine different multiresolution datasets. We present here an optimal interpolation approach to generate smoothed and gap-free time series of Landsat reflectance data. We fuse MODIS (moderate-resolution imaging spectroradiometer) and Landsat data globally using the Google Earth Engine (GEE) platform. The optimal interpolator exploits GEE ability to ingest large amounts of data (Landsat climatologies) and uses simple linear operations that scale easily in the cloud. The approach shows very good results in practice, as tested over five sites with different vegetation types and climatic characteristics in the contiguous US.
\end{abstract}
\section{Introduction}
\label{sec:intro}

Remote Sensing data constitute one of the most valuable sources of information for global land-cover mapping, crop monitoring, and phenological parameter inversion. Unfortunately, products derived from a single Earth observation sensor are hampered by a limited spatial, spectral or temporal resolutions~\cite{CampsValls11,gomez2015multimodal}. 
Fusion of remote sensing images with different characteristics is nowadays possible with different levels of sophistication through spatial-spectral image fusion, also known as {\em blending} or {\em downscaling}~\cite{Gao06,Hilker09,Amoros11,Villa13,Gevaert15}. 

Blending Landsat and MODIS (moderate-resolution imaging spectroradiometer), for instance, has enabled predicting daily surface reflectance~\cite{Gao06,Hilker09}, and to study NDVI (normalized difference vegetation index) evolution with denser time series~\cite{vanLeeuwen06}. On the one hand, MODIS provides daily global observations at medium resolution (250-1000m) which  allows tracking rapid land-cover changes and maximizes the chances of having cloud-free observations due its high temporal resolution, but limits its effectiveness for fine-scale environmental applications. On the other hand, the Landsat mission provides high spatial detail (30 m) ideal for monitoring human activity processes but the low temporal resolution (16 day revisit cycle) limits its use especially in areas with significant cloud occurrence. The creation of consistent Landsat time series requires to deal with data gaps, scene overlaps, sensor malfunctioning (Landsat 7), and inherent satellite data noise (due to sensor noise, partial pixel cloud and atmospheric contamination, etc.) \cite{robinson2017dynamic}.
Downscaling Landsat and MODIS yields improved spatial but also enhanced spectral products, thus allowing to compute indices beyond NIR (near infrared) : indices in the SWIR (short wave infrared) range enable more accurate vegetation studies, moisture assessment and standing water identification~\cite{Jarihani14}. 

Different approaches can be found in the literature to cope with these problems~\cite{cai2017performance}. Most of them use gap filling and smoothing methods which rely on temporal, spatial interpolation or both \cite{moreno2014noise}. Another kind of approaches use multi-sensor data fusion to reduce the constraints of single sensor remote sensing, among these, probably the spatial and temporal adaptive reflectance fusion model (STARFM) is the most common data fusion algorithm for MODIS and Landsat data \cite{gao2006blending}. An important drawback of the STARFM algorithm is that it needs complex spatial operations to account for inhomogeneities within coarser MODIS pixels. This operation is computationally very expensive and prevents this algorithm to be implemented globally \cite{rao2015improved}. A simpler and very fast approach was proposed by \cite{hwang2011downscaling}, their method used a linear relationship between FAPAR (fraction of photosynthetically active radiation) and NDVI, and allowed to downscale MODIS coarser resolution data using Landsat TM data. A completely different method to fuse MODIS and Landsat was proposed by \cite{sedano2014kalman}, where a Kalman filter was implemented to integrate medium and moderate resolution imagery using spatio-temporal information.

In this work, we show preliminary results of an operational approach \cite{daley1993atmospheric} to fuse MODIS and Landsat globally using the Google Earth Engine (GEE) platform. To do so, we have exploited the flexibility of an {\em optimal interpolator} along with the computational power of GEE to provide filtered and gap-filled time series of Landsat reflectance estimates using MODIS data, a Landsat ten-year climatology, and actual Landsat reflectance data.

The remainder of the paper is organized as follows. Section \S2 details the data collected
and processed in this study, as well as the experimental setup. Section \S3 introduces the methodology proposed for optimal interpolation and gap filling. Section \S4 gives the experimental results. Finally, Section \S5 concludes the paper and outlines the further work.

\section{Data collection and setup}
\label{sec:data}

We used a large quantity of data derived from both MODIS and Landsat satellites collected over many years, as detailed in Table~\ref{tab:RSdatadescription}. For illustrative purposes, in this preliminary study, we have used the red and infrared bands of MODIS (Band 1, 0.62-0.67 $\mu m$ and Band 2,  0.84-0.88 $\mu m$) and Landsat (Band 3, 0.63-0.69 $\mu m$ and Band 4, 0.77-0.90 $\mu m$) satellites.

\begin{table}[!h]
\centering
\caption{General description of the remote sensing products used in this work.}
\label{tab:RSdatadescription}
\begin{tabular}{ | l |  m{1.9cm} |  m{2.6 cm} |}
\hline
\hline
Product &  Period & Temporal/Spatial resolutions \\
\hline
\hline
Landsat 5 SR &  1999-2010  & 16 days / 30 m \\
\hline
Landsat 7 SR  &  1999-2010  & 16 days / 30 m \\
\hline
MOD9A1 (Terra) &  2010 & 8 days / 500 m \\
\hline
MYD9A1 (Aqua) &  2010 & 8 days / 500 m \\
\hline
\hline
\end{tabular}
\end{table}

To reduce computational needs and remove part of the noise we aggregated the data to a monthly temporal resolution. When multiple images were available for a given month, the median value  for each pixel was calculated. We have only considered data with the maximum quality level, according to the quality flag information of the corresponding remote sensing product.

In order to obtain typical monthly values of the considered spectral bands, we have computed a monthly climatology (median values for each month for all considered years) combining 10 years of data from Landsat 5 and Landsat 7 (1999-2009) satellites. To quantify the temporal variance of the climatology, we additionally computed the standard deviation for each month.

To exploit the better temporal resolution of the MODIS sensor, we have also included a simple linear fusion algorithm assuming that the Landsat and MODIS reflectances could be related linearly. This model is independent for each Landsat pixel, allowing to capture potentially in its parameters spatial heterogeneities within a coarser MODIS pixel. More sophisticated, eventually nonlinear, interactions will be considered in the future within our framework.

\section{Proposed methods}
\label{sec:methods}

For the data assimilation and sequential filtering problem, we consider the following specific model
\begin{gather}
\label{Super_Model}
\left\{
\begin{split}
 x_k&\sim \mathcal{N}(u_{k,1},P_{k,1}),\\
 u_{k,2}&=x_k+\xi_{k}, \\
y_k&=H x_k+v_k, \\
\end{split}
\right.
\end{gather}
where $k\in\mathbb{N}$ is the iteration index, $x_k\in \mathbb{R}$ is the variable of interest (which we desire to infer), $y_k\in \mathbb{R}$ is the observed measurement at the $k$-th iteration, $H \in \mathbb{R}$ is a known constant,  $\xi_k\sim \mathcal{N}(0,P_{k,2})$, and $v_k\sim \mathcal{N}(0,R)$ are Gaussian noise perturbations. Moreover, $u_{k,1}$ and $u_{k,2}$ are two known input signals: the Landsat climatology, $u_{k,1}$ and a  pixel-wise linear regression model to fuse MODIS and Landsat reflectances $u_{k,2}$.  The goal is to obtain a filtering solution $\widehat{x}_k$ given all the observations received, $y_1,\dots,y_k$, given the model in Eq. \eqref{Super_Model}. The optimal interpolation (OI) algorithm is a specific version of the Bayesian filtering methods (as the Kalman filter) for models of type  \eqref{Super_Model} \cite{daley1993atmospheric, sedano2014kalman}.
The prediction step of the OI algorithm consists in computing mean and variance of the predictive posterior considering the first two equations in model \eqref{Super_Model}, i.e.,
\begin{eqnarray}
\label{gausmix}
  x_k^-&=&u_{k,1}\dfrac{P_{k,2}}{P_{k,1}+P_{k,2}} + u_{k,2}\dfrac{P_{k,1}}{P_{k,1}+P_{k,2}}, \\
P_k^-&=& \left(\dfrac{1}{P_{k,1}} + \dfrac{1}{P_{k2}}\right)^{-1}
\end{eqnarray}
Then, the update step is given
\begin{equation}\label{statevector}
  \widehat{x}_k=x_k^- +K_k(y_k-H x_k^-),
\end{equation}
where the Kalman gain is
\begin{equation}\label{gain}
  K_k=\frac{H P_k^-}{H^2 P_k^- +R},
\end{equation}
Furthermore, the algorithm provides the posterior variance by
\begin{equation}\label{gain}
  P_k=(1- K_k H)P_k^-,
\end{equation}
representing the uncertainty in the estimation.

\section{Experimental results}
\label{sec:resuts}

\begin{figure}[t!]
  \centering
  \includegraphics[width=9cm]{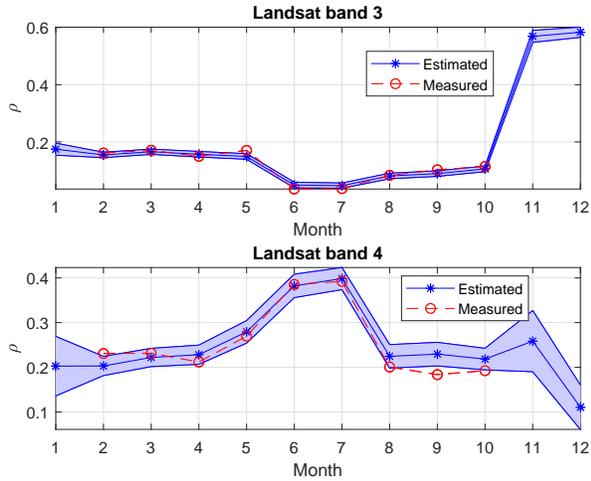}\\
  \caption{Results obtained in a crop land area in Montana (USA) in 2010.}\label{cropfig}
\end{figure}

\begin{figure}[t!]
  \centering
  \includegraphics[width=9cm]{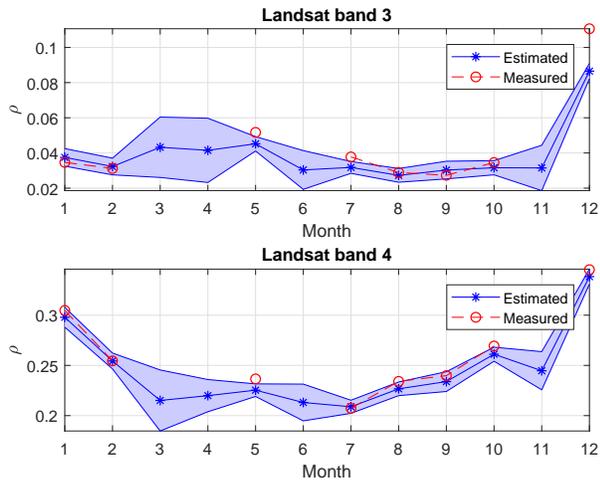}\\
  \caption{Results obtained in a forest site with high cloud occurrence in Seattle (USA) in 2010.}\label{forestfig}
\end{figure}

Figures \ref{cropfig}-\ref{forestfig} show two examples of application of the developed optimal filter: over a crop land area and over a forest location with high occurrence of clouds.  As can be seen, the estimated reflectances match adequately the temporal variation of the measured Landsat reflectance. Additionally, the proposed method seems to correctly recover the missing data due to cloud contamination preserving the expected gradual temporal evolution when no other factors like snow are affecting the reflectances (first and last months of the year). Because of the characteristics of the followed approach, we also provide confidence intervals of our estimates. These can be very useful for uncertainty propagation purposes in future applications. It is also important to note that, when no actual measurements were available, the optimal filter is automatically giving higher uncertainties in their estimates (see months three and four in figure \ref{forestfig} for example).

The optimal interpolator has been tested and evaluated in five different test sites (Table \ref{table:evaluation}) using a leave-one-out cross-validation approach. Sites $\sharp$1 and $\sharp$2 correspond to crop land areas in Montana, US; site $\sharp$3 corresponds to a cloudy evergreen forest area in Seattle, US; $\sharp$4 represents a deciduous forest in Montana, US; and site $\sharp$5 is located in an evergreen forest in Georgia, US.

As can be seen in the results, the followed approach produces low biased estimates for all sites in both considered bands. Errors are significantly higher in band three than in band four, this is explained by the fact that all sites considered correspond with vegetated areas. Vegetated areas are characterized by low reflectance responses (low signal to noise ratios) in the red band because of the chlorophyll absorption  and much higher reflectance in the near infrared bands (high signal to noise ratio). This spectral signature has been largely documented in the literature and it is the foundation of the most commonly used remote sensing spectral indices (NDVI, EVI, RDVI, etc) to monitor land vegetation globally.

\begin{table}[t!]
\small
\caption{Cross-validation results for all considered sites.}
\label{table:evaluation}
\begin{center}
\begin{tabular}{|l|c|c|c|c|}
\hline
\hline
{\bf Site}	 & {\bf ME}	 & {\bf RMSE} & {\bf MAE} & {\bf Mean $\rho$} \\
\hline
\hline
\multicolumn{5}{|c|}{Band 3} \\
\hline
1	 &  -0.004	 & 0.03	 & 0.02 & 0.10\\
2	& -0.04 & 0.10	 &0.06 & 0.15 \\
3    & 0.009	 & 0.03	 & 0.02 & 0.03\\
4	 &  -0.002	 & 0.019	 & 0.013 & 0.04 \\
5	 & 0.0014	 & 0.010	 & 0.008 & 0.04 \\
\hline
\multicolumn{5}{|c|}{Band 4} \\
\hline
1	 & -0.04	 & 0.08	 & 0.06 & 0.3\\
2	 & 0.05  & 0.08	 & 0.06 & 0. 3 \\
3	 & 0.03	 & 0.04	 & 0.03 & 0.2 \\
4    & 0.007	 & 0.016	 & 0.012  & 0.09\\
5	 & 0.03	 & 0.03	 & 0.03 & 0.18 \\
\hline
\end{tabular}
\end{center}
\end{table}

\section{Conlusions}
\label{sec:conclusions}

In the present paper, we have shown the proof of concept for a global approach to fuse MODIS and Landsat data using the GEE platform. The proposed optimal interpolation method is ideal to be implemented in GEE because it exploits GEE ability to process large amounts of data (Landsat climatologies) and uses simple linear operations that scale easily in the cloud (no costly spatial calculations are needed). The performance of the proposed methodology has been preliminary tested over five sites with different vegetation types and climatic characteristics located in the contiguous US. The quality of the estimated spectral profiles were consistent for all sites and both considered bands, although more tests in different areas need to be carried out in the future to better analyze the robustness of the followed approach.

Because of the nature of the proposed method, it could be easily extended to work for the rest of the bands of the Landsat sensor and could be adapted to use other sensors like Sentinel-2 or the Visible Infrared Imaging Radiometer Suite (VIIRS) for example. The presented methodology offers many opportunities to combine different remote sensing data sources to obtain consistent and continuous time series with their associated uncertainty estimates. The development of fusing, downscaling and gap filling methods in the GEE plattform constitutes a necessary stepping stone before more ambitious far-end goal of accurate, time-resolved, multi-decadal global map analysis~\cite{Fensholt09,Fensholt12,Hansen13} can be carried out with reliable input data.

\small
\bibliographystyle{IEEEbib}
\bibliography{refs,blender_simpler}

\end{document}